\documentclass[journal,twocolumn]{IEEEtran}

\ifCLASSINFOpdf
\else
\fi
\let\labelindent\relax

\usepackage[hidelinks]{hyperref}
\usepackage{graphicx}
\usepackage{adjustbox}
\usepackage[utf8]{inputenc}
\usepackage{booktabs}
\usepackage{amsmath}
\usepackage{amssymb}
\usepackage[numbers]{natbib}
\usepackage{paralist}
\usepackage{multirow}
\usepackage{xspace}
\usepackage{color}
\usepackage{xcolor}
\usepackage[graphicx]{realboxes}
\usepackage{ifthen}
\usepackage{url}
\usepackage{fancybox}
\usepackage{enumitem}
\usepackage{listings}
\usepackage{balance}
\usepackage{ifthen}
\usepackage{graphicx}
\usepackage{amsmath}
\usepackage[linesnumbered,boxruled]{algorithm2e}
\usepackage{algorithm2e}
\usepackage{tablefootnote}
\usepackage{comment}
\usepackage{float}
\usepackage{algorithmic}
\usepackage{dirtytalk}
\usepackage[tikz]{bclogo}
\usepackage{subcaption}
\usepackage{tcolorbox}
\usepackage{booktabs}
\usepackage{tikz}
\usetikzlibrary{plotmarks}
\usetikzlibrary{arrows,shapes,positioning}
\usetikzlibrary{decorations.markings}
\tikzstyle arrowstyle=[scale=1]
\tikzset{>=latex}

\usepackage{pgfplotstable}
\graphicspath{{pics/}}

\definecolor{codegreen}{rgb}{0,0.6,0}
\definecolor{codegray}{rgb}{0.5,0.5,0.5}
\definecolor{codepurple}{rgb}{0.58,0,0.82}
\definecolor{backcolour}{rgb}{0.95,0.95,0.92}

\lstdefinestyle{mystyle}{
	backgroundcolor=\color{backcolour},   
	commentstyle=\color{codegreen},
	keywordstyle=\color{magenta},
	numberstyle=\tiny\color{codegray},
	stringstyle=\color{codepurple},
	basicstyle=\footnotesize,
	breakatwhitespace=false,         
	breaklines=true,                 
	captionpos=b,                    
	keepspaces=true,                 
	numbers=left,                    
	numbersep=2pt,                  
	showspaces=false,                
	showstringspaces=false,
	showtabs=false,                  
	tabsize=2
}

\usepackage[english]{babel}
\usepackage{epigraph} 
\DeclareGraphicsExtensions{.pdf,.jpeg,.png}
\begin{document}

\newtheorem{theorem}{Definition}[section]
	\newcommand{\eg}{e.g.,}
	\newcommand{\ie}{i.e.,}	
	\renewcommand{\lstlistingname}{Listing}

\newcommand{\boxedtext}[1]{\fbox{\scriptsize\bfseries\textsf{#1}}}
\newcommand{\nota}[2]{
	\boxedtext{#1}
		{\small$\blacktriangleright$\emph{\textsl{#2}}$\blacktriangleleft$}
}

\newcommand\review[3]{\textcolor{red}{\sout{#1}} {\textcolor{blue}{#2}}{\todo{#3}}}

\newcommand{\Rqone}{RQ$_1$: \textit{To what extent do developers make contributions to the ecosystem?}}
\newcommand{\Rqtwo}{RQ$_2$: \textit{What kinds of contributions are made to the ecosystem?}}
\newcommand{\Rqthree}{RQ$_3$: \textit{What are the motivation to contribute to the  ecosystem?}}

\definecolor{beaublue}{rgb}{0.74, 0.83, 0.9} 
\definecolor{redbeau}{HTML}{cc5b5b}  
\definecolor{orabeau}{HTML}{ff9f79}

  \title{Ethical Considerations Towards Protestware}

\author{
\IEEEauthorblockN{Marc Cheong\IEEEauthorrefmark{2}, 
Raula Gaikovina Kula\IEEEauthorrefmark{1}, and
Christoph Treude\IEEEauthorrefmark{2}}\\
\IEEEauthorblockA{\IEEEauthorrefmark{2}University of Melbourne, Australia,
\IEEEauthorrefmark{1}Nara Institute of Science and Technology, Japan
\\
marc.cheong@unimelb.edu.au, christoph.treude@unimelb.edu.au, raula-k@is.naist.jp }
}

\maketitle

\begin{abstract}
A key drawback to using a {free and open source software (FOSS)} third-party library is the threat that it opens up a doorway to deliver malicious attacks.
In recent times, these threats have taken a new form, when the library maintainers themselves turn their {FOSS} libraries into protestware.
This is defined as software containing political messages delivered through these libraries, which can either be malicious or benign.
{
Since developers readily adopt these libraries into their software, much trust and responsibility are placed on the maintainers to ensure that the library does what it promises to do.}  
This paper takes a look into the possible scenarios where developers might consider turning their FOSS into protestware, using an ethico-philosophical lens.
Using different frameworks commonly used in AI ethics, we {extend the applications of the ethics of AI to the study of protestware, paying attention to its multi-perspectival nature, involving stakeholders -- or moral agents -- of varying degrees of involvement and participation, cognisant of the fact that ethical reasoning is complex}.
Additionally, we illustrate how an open-source maintainer's decision to protest is influenced by different stakeholders (\textit{viz.}, their membership in the FOSS community, their personal views, financial motivations, social status, and moral viewpoints), making protestware a multifaceted and intricate matter. {We also explore potential perspectives from the other stakeholders in the FOSS ecosystem, \textit{vis-à-vis} the negative and positive effects of protestware and consider their status as moral agents in the community.}
\end{abstract}

\epigraph{``When people feel they are not being heard, they may resort to different measures to get their message across. In the case of programmers, they have the unique ability to protest through their code.''}{Kula \& Treude (2022) \cite{kula2022war}}

\epigraph{``Consequently, he found himself confronted by two very different modes of action; the one concrete, immediate, but directed towards only one individual; and the other an action addressed to an end infinitely greater... but for that very reason ambiguous... He had to choose between those two. What could help him to choose?''}{ Sartre (1946)\\ \textit{Existentialism Is a Humanism} (trans.:  Mairet) \cite{SartreJean-Paul1948Eah}}

\section{Introduction}
In this article, we articulate the motivations behind maintainers who turn their {free and open source software} (FOSS) into protestware.
 {
There have been cases of self-sabotage with apolitical angles, such as for one's personal\footnote{\url{https://www.theregister.com/2016/03/23/npm\_left\_pad\_chaos/}}, economic\footnote{\url{https://github.com/zloirock/core-js/issues/548}}, or advertising\footnote{\url{ttps://www.zdnet.com/article/popular-javascript-library-starts-showing-ads-in-its-terminal/}} gain. However, in this paper, we focus on protestware, as it surpasses an individual's needs, and becomes an ethical issue that has larger societal ramifications. 
Protestware, especially in the context of FOSS, creates such a shift from traditional forms of protest, into the digital world that has the power to impact millions by ruining a single piece of code. 
}
Although ethics in computing is not new, the phenomenon of Protestware is unique, in that the power of responsibility is placed on individuals (i.e., sometimes a single library maintainer).

 {In this work, we carefully examine Protestware under the lens of three ethical frameworks (Duty, Consequentialism, and Principlism). For each framework,}
 we then explore the dilemmas that a library maintainer may face {vis-à-vis the other FOSS community members; and frame their actions from each lens, guided by lessons in the extant field of applied ethics, drawing upon its modern use in evaluating AI technologies and their humanistic impacts.}
We also discuss potential guidelines and larger ethical implications for the open source, industry, research, and education sectors{, while bearing in mind the multi-perspectival nature of ethical analysis which has both harmful and beneficial implications}.

\section{Background}
\label{sec:background}
\subsection{Context}
\label{sec:context}
In March 2022, the maintainer of {node-ipc,\footnote{\url{https://github.com/RIAEvangelist/node-ipc}}} a widely used software library, intentionally introduced a vulnerability into their code. If the code was run within Russia or Belarus, it would attempt to replace all files on the user's device with a heart emoji.\footnote{\url{https://techcrunch.com/2022/07/27/protestware-code-sabotage/}} This critical security flaw (i.e., CVE-2022-23812~\cite{Web:CVE-2022-23812}) highlights the trend of programmers intentionally sabotaging their code for political purposes, a practice known as ``\textit{protestware}''~\cite{kula2022war}.

The malicious code was intended to overwrite arbitrary files depending upon the geolocation of the user's IP address: in essence, attacking software in specific locations.
Specifically, the affected versions 10.1.1 and 10.1.2 of the library check whether the host machine has an IP address in Russia or Belarus, and if so, overwrites every file it could with a heart symbol. 
Version 10.1.3 was released soon after without this destructive functionality, while Versions 10.1.1 and 10.1.2 were removed from the NPM registry.

 Responses from the community varied, including frustrations that led to insightful discussions.
One example from a contributor on the GitHub Discussions channel is shown below~\cite{Web:discussion}:
\begin{quote}
\textit{I'm very happy to see that the principles and character of many in tech (FOSS especially) remain clear enough to recognize how completely wrong this was. Of course, if the marketplace of current things keeps hammering away at this, it will benefit a small number of corporate giants (misplaced trust/safety). I hope we all start seeing these patterns as we grapple with a general blurring of lines between tools for marketing and weaponry.
It's essential to ask: what's the outcome and who benefits? I like to ask the faux ideologues ``who agrees with you?'' ``Isn't it strange how well aligned you are with a small number of very visible, influential, and powerful organizations?'' ``What's the fight and who is on which side, again?''
It's about competency, not power. Power feeds and is fueled by egocentrism (plainly, weak vanity). Competency comes from discovering your natural gifts and applying them.} (sic.)
\end{quote}

Another user from that GitHub Discussion explained how this affected the  {open source community} \cite{Web:discussion}:
\begin{quote}
 \textit{The trust factor of open source, which was based on goodwill of the developers is now practically gone, and now, more and more people are realizing that one day, their library/application can possibly be exploited to do/say whatever some random dev on the internet thought was `the right thing to do'.} (sic.)
\end{quote}
The maintainer in question defended his module on GitHub, saying that \textit{``this is all public, documented, licensed and open source''} \cite{kula2022war}. Earlier, there were more than 20 issues flagged against node-ipc about its behavior.
Some of the comments referred to the creation as \textit{``protestware, while others might call it malware''}  \cite{kula2022war}. 

We present another case where the protestware does not have malicious intent, but aims at increasing awareness.
The same maintainer of the node-ipc library then created the peacenotwar {library.\footnote{\url{https://github.com/RIAEvangelist/peacenotwar}}}
As explained by the maintainer, it serves as a non-violent protest against Russia's aggression.
Instead of malicious deletion of files, the module adds a message of peace on users' desktops~\cite{Web:peacecommit}.
The maintainer was quoted in the README file:
\begin{quote}
    \textit{I pledge that this module, to the best of my knowledge and skills, does not do any damage to anyone's data. If you do not like what this module does, please just lock your dependencies to any of my work or other's which includes this module, to a version you have code reviewed and deemed acceptable for your needs. Also, please code-review your other modules for vulnerabilities.}
\end{quote}

\subsection{Characterization}
\label{sec:characterization}

 {Protestware represents a paradigm shift from traditional forms of protest, such as strikes or demonstrations. Programmers have the unique power to embed their dissent directly within code. A small tweak to a widely-used software application can spread a protest message globally at an unprecedented speed. 
 Whereas traditional protests usually need mass mobilisation to be effective, a single individual can make a significant impact with protestware. Traditional protests are typically confined to specific legal jurisdictions, but protestware's cross-border nature makes its legal status ambiguous.

The fundamental ethical principles and frameworks guiding activists remain consistent across both protestware and traditional forms of protest. In this paper, we discuss these principles and frameworks specifically in the context of this emerging form of activism: protestware.}

Protestware can take  {two} forms  {(based on~\cite{kula2022war})}: 

\begin{description}
\item[malignant protestware] which intentionally damages or takes control of a user's device without their knowledge or consent;  {and}
\item[benign protestware] which raises awareness about a social or political issue without causing harm (e.g., changes to license files\footnote{\url{https://github.com/terraform-aws-modules/terraform-aws-ec2-instance/commit/6867788411a202b61187f9935e9eaa72a18f0bbe}}).
\end{description}

{These two forms of protestware can manifest in various ways, such as project documentation (e.g., README banner), communication (e.g., log messages), environment (e.g., injected code on target machines), or output (e.g., file deletions). 
In this study, we focus on protestware manifested in the environment (i.e., manifested in the source code), which, even if benign, tends to draw more attention of the user}.
Protestware can have a wide-ranging impact on numerous stakeholders, including other contributors to the same project, direct and indirect users of the project, and even the entire open-source community and its newcomers.

 {On one hand, the rise of protestware raises the question of whether it is ethical to intentionally worsen something to make a point. On the other hand, protestware draws attention to current issues with the potential for social change.} 
This issue is particularly significant in software ecosystems where code is frequently reused~\cite{zahan2022weak}, as these ecosystems rely on the trust and reliability of the code being used. If a programmer introduces protestware into their code, it can introduce a ripple effect: compromising the security and stability of the software for all users who reuse that code, potentially affecting individuals and businesses that rely on the software, as well as further dependencies.

\subsection{Protestware: Beyond Computing Ethics?}
The literature for computing and AI ethics is rapidly growing, predominantly on sociotechnical systems and ``artificial moral agents" \cite{Zoshak2021} such as algorithmic recommender systems, self-driving cars, algorithmic policing and law enforcement, and AI-based person recognition.

 {In the same way that `traditional' AI ethics adapts and contextualises applied ethics \cite{Zoshak2021, Muller2020-wp} to AI systems, protestware ethics similarly extends `traditional' AI ethics by recognizing the various stakeholders in the open source ecosystem. For starters, AI ethics issues typically focus on major actors such as technology providers or adopters \cite{coghlan}.}
What sets our proposed analysis of protestware apart from existing computing ethics studies is that the stakeholders involved have a more personal, direct, and explicit involvement. To illustrate: consider a company (\textit{BigCorp}) whose self-driving car who has injured a pedestrian: it is hard to see which individual programmer or engineer is responsible for the injury, based on the `diffusion' of responsibility through \textit{BigCorp}'s organisational structure.\footnote{ {We recognise that, in some instances, this `diffusion' resembles \textit{collective actions} such as blackouts in protest of the 2012 Stop Online Piracy Act (\url{https://en.wikipedia.org/wiki/Protests_against_SOPA_and_PIPA}). However, we emphasise the dynamic nature of protestware in the FOSS ecosystem, which can involve different levels of `diffusion' and collective action, \textit{per se}.}} Contrast this with an individual volunteer (\textit{Violet}), of an open source software library, who exhibits a form of activism by adding a few lines of code to prevent her software library from working in certain regions. 

In order to think about new guidelines for deciding ethical actions and consequences for \textit{Violet}, we will need to briefly explore the landscape of applied ethics.

\section{Ethics: A Primer}

\label{sec:ethics}
There exist various ethical theories, each with their own pros and cons \cite{Rachels2015}, which sometimes even conflict with each other in terms of their application and evaluation. 
Ethics, simply put, is about doing the ``\textit{right}'' things. However, there is no clear answer to what makes things ``right'' or ``moral''. Enter \textit{applied ethics} -- the application of {``one moral theory or [an]other} ... upon the applied ethics problem at hand, in the hopes of producing a resolution'' \cite{sep-theory-bioethics}. 

The field of medicine was one of the forerunners of this (e.g., by asking, \textit{Using common principles of medical ethics, how do we do no harm to patients under our care?}). This idea was quickly adapted into --- and gained traction within --- research in AI and computing in recent years (e.g., by similarly asking, \textit{How do self-driving cars do no harm to pedestrians and drivers we are responsible for, if those same principles of medical ethics are adopted and adapted?)}. 
Philosophers from Immanuel Kant and Jeremy Bentham in the 18th century, to Tom Beauchamp and James Childress in the 20th century, have proposed different approaches to the question of how to do so. Herein, we discuss three popular approaches, taking a leaf from the current state of computing ethics \cite{Zoshak2021}.

\subsection{Duty Ethics}
The ethical theory of \textit{deontology} -- or `duty ethics' -- basically posits that a moral agent\footnote{Simply put, a `moral agent' is usually a person who has agency to decide how to act.} has a ``sense of duty'' or requirement to do the right thing, which guides their actions. Immanuel Kant's `Categorical Imperative' offers a beautiful application of this, as summarised succinctly by Rachels: ``When you are thinking about doing something, ask what rule you would be following if you actually did it... Then ask whether you would be willing for your [rule]... to become a universal law'' \cite{Rachels2015}. In our example for \textit{Violet}, she ought to think about what would happen if \textit{all} programmers did the way she did, with no exception: would she be able to live with the consequences on a universal level? We could quickly see that there are exceptions to this: what happens if, say, by following her hypothetical universal law, one programmer's actions end up disabling life support machines in hospitals?

Following this maxim, 
hypothetically, if all 2,215,398 packages in the npmjs ecosystem decided to inject  {protestware} code into their systems, it would  {affect all libraries in the ecosystem -- with differing levels of severity, impact, and time needed to revert changes, based on the nature of their payloads --} as well as all the potential applications that adopt the libraries into the ecosystem. 
Since JavaScript is the top-ranked language on GitHub, this is a significant portion of GitHub projects as well.
Although her intent was to protest against persons or groups, this would disqualify the code from being {safely} distributed.

Beyond \textit{moral} harm, protestware will also disqualify the {FOSS} license, which has implications in the existing IT infrastructure.
For instance, a market report in 2019\footnote{https://www.fortunebusinessinsights.com/open-source-services-market-106469} shows that FOSS IT infrastructure is used by 53\% of organisations, 43\% integrations, and 42\% in digital transformation.  
While we recognise the possibility of ``ethical" licenses such as the DoNoHarm license and others\footnote{See e.g., \url{https://ethicalsource.dev/licenses/} and \url{https://github.com/raisely/NoHarm} for examples.}, we would like to underscore that duty ethics \textit{a la} Kant is a moral framework that reasons in terms of how ``a rational being [who] decides to treat people in a certain way... decrees that this is the way \textit{people} [\textit{i.e., as a whole}] are to be treated'' \cite{Rachels2015} (emphases ours).
\begin{tcolorbox}
\textbf{Duty Ethics:} An example of applying the Kantian Categorical Imperative to the dilemma would be, e.g., \textit{What if all programmers regard malicious code injection as ethically imperative?} One can argue that the library will ultimately violate the tenets of FOSS, leading to a ban on the library. Hence, by implication, malicious code injection is not ethical. {However, in the same vein, one could argue that, under the right circumstances, if all programmers added a benign notice that affirms solidarity with a particular social cause, there might be a momentum for large-scale social change}.
\end{tcolorbox}

\subsection{Consequentialist ethics}
Next, another ethical theory which appeals to many computing professionals, due to its `mathematical' nature, is \textit{utilitarianism} - under the broad banner of \textit{consequentialist} ethics. Simply put, it is the consequences that are to be the yardstick by which to measure one's initial actions; and it is one's responsibility to maximise the overall happiness (or `utiles' if you wish, mathematically speaking) across all stakeholders, and to minimise any unhappiness \cite{Rachels2015}. First proposed by philosophers such as Jeremy Bentham, it is thus easy to understand the attractiveness of this theory, as it boils down to an optimisation problem of `utiles'. However, translating it into practice is much harder than it looks on the surface: for starters, how could Violet measure the overall net gain of her activism? When the `utiles' are divided across the entire population, does an individual merely get an infinitesmal $+0.0001$? Also, who gets to be the arbiter of the quantity of the individual utiles? Most crucially, how would she quantify the overall net loss of unintended consequences? (Again with our life support machine example from before: is the unintended death of a person worth $-10,000$? $-100,000$? Some might decide that is is simply \textit{unquantifiable}\footnote{ {The \textit{value of a human life} is a nuanced topic, with different philosophical considerations and perspectives. Further explication is provided in ethics textbooks, such as \cite{Rachels2015}.}}!)

In terms of consequences, the risk of losing any of the users of the library, potential community of contributors, and also their standing in the FOSS community, are all net losses that need to be quantified. 

 {Unfortunately, there is no clear answer: and as is the case of many ethical evaluations, what might be an acceptable solution (or tolerable trade-off) in one case might not be for another. For starters, benign protestware might only require, say, an hour for maintainers to revert the changes to the last stable version (and nothing more); whereas malignant protestware would require substantially more time to undo the damage caused (i.e., a large difference in negative utiles levied on different members of the community).}

\begin{tcolorbox}
\textbf{Consequentialist Ethics:}
An example of a utilitarian's reasoning would take the form of, e.g., \textit{Potentially risking the ban of the library might be a bigger consequence (higher nett negative `utiles') than trying to target a subset of users to send a message (smaller nett positive `utiles'), leading to an overall negative in the balance of probabilities.} 
To put this reasoning in plain language: although the damage might have short term benefits, the long term effects and consequences are much larger. {A counterargument can also be made if the potential advantages -- again, e.g., lasting social change -- outweighs, e.g., a cosmetic change to a UI with no lasting repercussions.}
\end{tcolorbox}

\subsection{Principlism}
A more pragmatic approach will be to follow a set of fixed principles, in the namesake philosophical framework of \textit{principlism}. Here, we draw inspiration from Beauchamp and Childress' landmark \textit{Principles of Biomedical Ethics}, which posits four key principles: respect for autonomy; nonmaleficence (doing no harm); beneficence (doing good); and justice \cite{Beauchamp1994}. All these principles have been in use in biomedical ethics, and have seen promise in evaluating issues in technological ethics{, such as in deployment in AI-based solutions in education \cite{coghlan}}.

\begin{tcolorbox}
    \textbf{Principle 1 --- Respect for autonomy:} This principle involves respecting the freedom and autonomy between users, maintainer, and the broader FOSS community. This lies at the heart of FOSS, per e.g., the Free Software Foundation \cite{GNU}. 
\end{tcolorbox}

\textbf{Respect for autonomy:} In considering this principle, we have identified at least four parties involved: the users of a library; the maintainer; the contributor; and finally the ecosystem as a community of FOSS developers.
Since FOSS is driven by volunteer contributions, would curtailing any party's autonomy -- from denying them use of the software, to, say, causing them to work on weekends in order to revert changes to repositories or fix production code {(again, depending on severity and malignancy of protestware)} -- impact their freedom, or worse, change any of the parties' motivations? Overall considerations will involve respecting the freedom of choice for all parties involved.

\begin{tcolorbox}
    \textbf{Principle 2 --- Nonmaleficence:} This principle requires that harm \textit{should not} be caused to stakeholders. This includes indirect harm caused to others in the FOSS community based on their ``assumed belief, group membership, or behavior" \cite{coghlan}.
\end{tcolorbox}

\textbf{Nonmaleficence:} This, we reason, would only apply to the benign form of protestware. To recap, these forms will not cause undue impact, when compared to their malignant counterparts which may lead to security vulnerabilities. 
Critical questions when evaluating this principle includes, say, \textit{How does a protester manage to send a message, while not causing harm to any of the stakeholders.} 
There are examples of README placements of protestware\footnote{\url{https://github.com/vshymanskyy/StandWithUkraine}}: while delicate, this is a strategy to use documentation and communication channels, as opposed to modification of the actual code.

\begin{tcolorbox}
    \textbf{Principle 3 --- Beneficence:} This principle stipulates that consequence of an action should result in good or benefit. Note that this does not just consider the provision of benefits, but also, ``balancing benefits against risks and costs" \cite{Beauchamp1994}.
\end{tcolorbox}

\textbf{Beneficence:} As we have seen in our treatise on utilitarianism, the benefits of protestware could be difficult to quantify at this stage. Nonetheless, an argument can be made for the benefits of fund-raising efforts such as sponsors or ad campaigns to create awareness on the political situation. 
This could be in the form of incentives to users, and to contributors of the source code.
For historical context, however, a common incentive in sabotaging-one's-own-code was to help sponsor struggling maintainers who wanted to protest against their software being used by large corporations.\footnote{https://www.independent.co.uk/tech/developer-sabotages-code-protest-github-colors-faker-b1990161.html}

\begin{tcolorbox}
    \textbf{Principle 4 --- Justice:} The effects of actions must be just and fair in terms of the distribution of ``benefits, risks, and costs" \cite{Beauchamp1994}.
\end{tcolorbox}

\textbf{Justice:}
Finally, the argument for justice requires not just fairness of outcomes, but also its \textit{distribution}, which includes the risks and costs \cite{Beauchamp1994}. 
In a historical context, it is prudent to consider the protest against usage of FOSS by the industry (which precedes protestware) which some developers consider to be unfair, raised in \textit{Beneficence} above. The original \textit{raison d'\^{e}tre} involves getting justice against the actual corporations that use the library. However, at the heart of this principle, other users and stakeholders may be disproportionately affected by this maneuver which leads it into question. 

\vspace{8pt}
Think of these as valuable \textit{heuristics} for which we could evaluate the fitness of an action we are taking. {As seen earlier, particularly in the quote in our epigraph, philosophers recognise the difficulty of doing the right thing or knowing what in fact \textit{is} the right thing \cite{Rachels2015} which is the difficulty faced by a moral agent when confronted with a difficult choice. Despite} our best intentions, we might run into trouble fairly quickly. For example, what do we do when we do not satisfy all the listed heuristics; Violet might (indirectly) cause harm and deny autonomy to innocent users who are affected by her code destruction, while still advocating for perceived justice and beneficence, based on the cause of her activism? {The next section will provide us with some guidance as to how to proceed.}

\section{ {Guidelines for} Promoting Ethical Responsibility}
\label{sec:guidelines}

Assessing the ethical implications of an open source maintainer's decision to convert their software into protestware is a multifaceted and intricate matter. For starters, the ethical frameworks may not agree with each other: as we have seen, the Kantian Categorical Imperative (duty ethics) might be a straightforward ``don't do it".\footnote{{It is important to note that the focus of the paper is separate from the assumptions and trust of FOSS components.}}

Drawing upon principlism, however, the nuance is visible. From the standpoint of autonomy, maintainers possess the right to make choices regarding \textit{their own} creations. However, this decision may conflict with \textit{end users'} autonomy, as it could restrict their ability to utilize the software without unforeseen --- or disastrous --- consequences. The principles of beneficence and non-maleficence are also crucial to consider: although protestware might advance a greater good by raising awareness or advocating change, it could simultaneously inflict harm upon users who depend on the software for essential functions (again, with our life support machine example raised several times before). The justice principle underlines the importance of examining the fairness and equitability of deploying protestware as a protest method. Hence, it is vital to evaluate whether such actions disproportionately impact specific groups (including, e.g., time and effort to rectify deleterious code behavior) or inadvertently create disparities in software resource access. Bearing these ethical principles in mind, determining the suitability of transforming FOSS into protestware necessitates a thorough analysis of the potential benefits, drawbacks, and broader societal ramifications of such a decision.

In this section, we present various initiatives which can be implemented to promote a more balanced perspective on protestware --- with an emphasis on its potential risks --- and encouraging maintainers to consider alternative methods for expressing their concerns. We also suggest directions for future work, including examining the role of FOSS governance, policy, and the `social license to operate', before finally identifying ways to protect users from protestware threats.

\paragraph{Responsibility in the FOSS Community}
To minimize the moral dilemmas (and concrete implications) associated with protestware, maintainers should be encouraged to prioritize the needs of their users and contribute to the greater good of the {FOSS community (as a start), with a view towards the \textit{common good} in the broader community of adopters and users.} Cultivating a sense of community and fostering strong relationships with stakeholders --- from end users to fellow developers --- can help maintainers understand the potential consequences of their actions and work together to develop ethical guidelines. Establishing communication channels and fora for discussion allows maintainers to express their concerns without resorting to protestware in the first instance, ensuring the well-being of both the community and its users. Future work should investigate how the community can cultivate a culture of care\footnote{Other frameworks of ethics, including care ethics, are also considered in current research into technology.} and moral responsibility and explore existing channels that allow maintainers to voice their protests in a non-jeopardizing way.

\paragraph{Safeguards against Protestware}
Future work could also explore the use of machine learning techniques, particularly natural language processing and code analysis algorithms, to detect early indicators of potential protestware in software repositories. Such techniques, in the same vein as, e.g., automated auditing for privacy and security issues, could better equip end users or developers to protect themselves from potential threats, ultimately enhancing the overall security and trustworthiness of the open-source ecosystem.

\paragraph{Enabling Healthy Channels for Protest}
It is important for maintainers to feel that their concerns are being addressed through appropriate channels. Good governance and a fair system of representation can alleviate the need for protestware by offering maintainers alternative avenues to voice their opinions and effect change within the FOSS community. Future research should explore how existing channels for maintainers to voice their protests or grievances can be improved or expanded. {Beyond the avenues of open dialogue and representation, licensing plays a pivotal role in shaping the use and dissemination of open-source software. The choice of license can serve as a proactive mechanism for maintainers and contributors who wish to place constraints on the use of their software. For example, choosing the Affero General Public License (AGPL) may dissuade certain corporate entities from using the software due to its strong copyleft provisions. Alternatively, maintainers can adopt licenses that impose direct usage restrictions to specific entities or purposes, although this would deviate from the standard open source definitions. Such licensing strategies can provide maintainers with a means to set boundaries and expectations, reducing the perceived need for more confrontational approaches like protestware.}

\paragraph{Education of Ethical Responsibility}
While this article is a good first step to foster awareness, ethics educational programs, focusing on ethical responsibility and social impact, can help maintainers better understand the potential consequences of using protestware and foster a more nuanced perspective on this issue. By providing guidance on ethical decision making and highlighting alternative methods for expressing concerns, ethics education can play a vital role in promoting a more balanced and responsible approach to protestware within the FOSS community. Future work should examine why some developers might feel far removed from ethical considerations and how educational programs can effectively address this issue.

\section{ {Implications with Future Directions}}
Taking into account the positives and negatives with the emergence of protestware, we depict the diverse ethical challenges for various stakeholders {, each with their own perspectives and priorities.}
In the following paragraphs, we discuss the implications for different stakeholders, emphasizing the importance of awareness and providing examples of ethical frameworks that could be applied in these situations.

\textit{Maintainers of the code}, when confronted with protestware, play a pivotal role in shaping the FOSS landscape. It is crucial that they clearly communicate the intended use and restrictions of their software in the documentation and/or terms of service while adhering to applicable laws and regulations. As an example, duty ethics highlights the importance of maintainers' moral obligations, such as transparency, honesty, and legal compliance.

\textit{Contributors who contribute but are not maintainers of the code}, as creators of public goods, must exercise responsibility, consideration, and ethics in their actions, particularly when engaging with FOSS projects. It is essential for contributors to be aware that any project they contribute to could potentially transform into protestware, and they may have limited control over the project's direction despite their contributions. Consequentialist ethics, for instance, emphasize the need to assess the potential consequences of one's actions, urging contributors to be mindful of the projects they engage with, maximizing positive outcomes, and minimizing potential harm. Awareness of protestware's implications and the potential risks of contributing to projects that may adopt such a stance is key to making informed decisions.

\textit{Newcomers} to the FOSS community should familiarize themselves with its principles, values, and ethical implications of protestware. FOSS relies on community-driven efforts, making it essential for newcomers to be respectful, collaborative, and helpful. Awareness of the presence of protestware and potential consequences is critical. As an example, the principle of autonomy in biomedical ethics underscores the importance of respecting the choices and values of other members of the community.

\textit{End users and industry} must be vigilant about the potential risks and benefits associated with using FOSS tools, acknowledging that any project could potentially transform into protestware. A key ethical implication for end users and industry is understanding that an FOSS project's maintainer might have different ethical priorities, which in extreme cases could lead to the creation of protestware. The risk is especially pronounced if end users or industry already rely on the software, and then it turns into protestware. Industry, in particular, must be cautious when using FOSS in a professional setting, ensuring compliance with company policies and regulations. By assessing risks, addressing potential security vulnerabilities, and contributing back to projects when possible, the industry can apply ethical principles like beneficence and non-maleficence from biomedical ethics, as an example, to promote positive outcomes and minimize harm to stakeholders. Awareness of the potential impact of protestware and the varying ethical priorities of project maintainers is crucial for responsible decision-making within the industry.

\textit{Educators} have a responsibility to discuss the implications and risks of FOSS, including protestware and its legal and ethical consequences. Duty ethics, as an example, encourages educators to impart knowledge of moral obligations and duties as software creators and users. Guided by consequentialist ethics and biomedical principles, students should be encouraged to consider the long-term impacts and ethical implications of their work, particularly with regard to protestware. The promotion of awareness among students is a key educational goal.

\textit{Researchers} play a vital role in identifying and mitigating protestware risks by developing methods to analyze code for vulnerabilities or detect patterns in malicious software behavior. By investigating the pros and cons of protestware as a political protest tool, researchers can apply ethical frameworks such as consequentialist ethics and biomedical principles to balance potential benefits and harms while considering justice and fairness within the FOSS community.

\bibliographystyle{IEEEtran}
\bibliography{filtered_ref}

\end{document}